
\def\PRep{Phys.\ Rep.\ }

\def\ZP{Z.\ Phys.\ }
\def\NP{Nucl.\ Phys.\ }
\def\PL{Phys.\ Lett.\ }
\def\T{\perp}
\def\pT{p_\T}
\def\qT{q_\T}
\def\rT{r_\T}

\def\Pol{{\cal P}}
\def\vpT{\vec p_\T}
\def\vqT{\vec q_\T}
\def\vhT{\vec h_\T}
\def\vrT{\vec r_\T}
\def\vPol{\vec{\cal P}}
\def\Aqq{{\cal A}_{q \to q'}}
\def\up{\!\!\uparrow\!\!}
\magnification\magstep 1
\font\bbf=cmbx10 scaled\magstep2
\pageno=0
\footline{\ifnum\pageno=0 \hfil \else \hss\tenrm\folio\hss\fi}

\hskip 9.5cm {\bf HEN-377(1994)}
\vskip 2.5cm

\centerline{\bbf Single spin asymmetry of vector meson production}
\vskip 0.2cm
\centerline{\bbf as a probe of asymmetry of parton scattering}
\bigskip
\bigskip

\centerline{{\bf J. Czy\.zewski}\footnote*{On leave from
\it  Institute of Physics, Jagellonian University,
ul.~Reymonta 4, PL-30-059 Krak\'ow, Poland}}
\centerline{\it Institute of High-Energy Physics, University of Nijmegen,}
\centerline{\it Toer\-nooi\-veld~1, NL-6525~ED Nijmegen, The Netherlands}
\medskip

\vskip 2cm
\centerline{\bf Abstract}
\smallskip

Azimuthal asymmetry of vector-meson production in single-transversely polarized
proton-proton collisions ($p\up p$) is calculated in the string model of
particle production.  The asymmetry is generated only during fragmentation of a
high-energy quark into hadrons.  The obtained asymmetry of the $\rho^\pm$ is
opposite in sign to that of $\pi^\pm$ mesons.  On the other hand, an asymmetry
appearing during parton scattering would contribute with the same sign to that
of vector and pseudoscalar mesons.  Thus, a combined measurement of both
can be used to estimate the contribution to the asymmetry from parton
scattering.

\vfill\eject

\noindent
{\bf Introduction}

\smallskip
\noindent
Measurement of single transverse-spin azimuthal asymmetries of particle
production in $p\up p$ collisions can provide information about transversely
polarized quark distributions in the proton [1].  However, to extract
information about the latter from the experimental data, one has to know the
mechanism generating the asymmetry.  It has been shown recently [2] that the
asymmetries of pion production measured by the E704 experiment in the forward
region can be explained by the Collins effect [3], {\it i.e.\ }the asymmetry
appearing at the level of the fragmentation of a transversely polarized
high-energy quark into hadrons.  In Ref.~[2] the string model was used to
describe the fragmentation, and the polarization effects were parametrized as
prescribed by the Lund model [4].  Positive asymmetry was obtained for $\pi^+$
and $\pi^0$ production and negative for $\pi^-$, resulting from upward
(downward) polarizations of the $u$ ($d$) valence qark in the proton polarized
upwards.

Different mechanisms leading to the azimuthal asymmetry are however possible.\
Szwed has shown [5,6] that the asymmetry can appear due to double gluon
exchange
during scattering of a quark on an external strong field.  This asymmetry
vanishes at a sufficiently high energy due to chiral symmetry and the resulting
helicity conservation.  Nevertheless, it can be measurable at finite energies.

In this note we present a method of distinguishing between the two scenarios.
We
calculate the asymmetry for vector mesons $\rho^\pm$ along the lines of
Ref.~[2].  This asymmetry appears to be opposite in sign to that of pions.  On
the contrary, if the asymmetry were determined at the hard or semi-hard
scattering, like in the approach of Szwed, then the asymmetries of pseudoscalar
(PS) and vector (V) mesons would not differ strongly.

To be more precise, we consider the reaction:
$$ p\!\uparrow + \, p \rightarrow h + X,
\eqno(1)$$
where $\uparrow$ refers to the projectile proton polarized vertically upwards
(parallel to the $\hat y$ axis) if the beam momentum points in the $\hat z$
direction.  The produced hadron $h$ carries the fraction $x_F = 2 p_z /
\sqrt{s}$ of the center-of-mass longitudinal momentum and the transverse
momentum $\vec \pT$.  In measurements of the asymmetry, the polarized
cross-section $d\sigma_\uparrow$ is assumed to behave as

$$ d\sigma_\uparrow(x_F, \vec\pT)= d\sigma(x_F,\pT)
   [1 + A_N(x_F,\pT) \cos(\phi)],
\eqno(2)$$
where $d\sigma$ denotes the unpolarized cross-section.  This defines the
asymmetry $A_N$.  $\phi$ is the azimuthal angle of the transverse momentum
$\vec\pT$ of the hadron, measured with respect to the $\hat x$ axis.

If one assumes factorization, the cross-section for the reaction (1) can be
written as a convolution of the parton distribution functions $G_{q/p}(x,\vqT)$
and $G_{r/p}(y,\vrT)$ in the projectile and the target protons, the
cross-section $d\hat\sigma$ for the parton scattering $q+r \to q'+r'$ and the
fragmentation function $D_{h/q'}$ of the quark $q'$:

$$
{d\sigma_\uparrow\over d^3\vec p} =
\int dx\, d^2\qT\ G_{q/p}(x,\vqT) \int dy\, d^2 \rT\ G_{r/p}(y,\vrT)\ \times
$$

$$
\int d\cos\hat\theta\, d\hat\varphi
\,{d\hat\sigma(\vPol_q,\hat\theta,\hat\varphi)
\over d\hat\Omega}
\int dz\, d^2 \vhT
D_{h/q'} (\vPol_{q'},z,\vhT) \, \delta^3(\vec p-z\vec q\,'-\vhT).
\eqno(3)$$
The summation over the flavours of $q$, $q'$, $r$ and $r'$ is understood.
$\vPol_q$ is the polarization vector of the quark $q$. Its magnitude is
defined by the transversity distribution [7]:

$$
\Pol_q = {\Delta_\T G_{q/p} \over G_{q/p}} =
         {G_{q\uparrow/p\uparrow} - G_{q\downarrow/p\uparrow} \over G_{q/p}}
\eqno(4)$$
The spin effects have been included by the dependence of $D_{h/q}$ on
$\vPol_{q'}$ (Collins effect) and of $d\hat\sigma$ on $\vPol_q$ (Szwed effect).
The polarization $\vPol_{q'}$ of the scattered quark $q'$ does not differ
strongly from $\vPol_q$ since the depolarization factor is close to unity at
typical small scattering angles $\hat\theta$ [3].

Theoretically, two extreme cases could be defined:

\item{\it a)} the asymmetry appears only in the fragmentation function as
calculated in Ref.~[2].  It manifests itself in
a dependence of the fragmentation function $D_{h/q'}$ on the azimuthal angle of
the transverse momentum $\vec h_\T$ of the hadron $h$.  This is the case of the
high-energy limit since any asymmetry appearing at the parton level (in $d\hat
\sigma$) must vanish in that region due to chiral symmetry.  In this case
the asymmetry can depend on whether $h$ is a PS or a V meson.

\item{\it b)} the asymmetry appears only at the parton level {\it i.e.}  $d\hat
\sigma$ depends on the azimuth $\hat\varphi$ of the parton scattering plane.
It
has been shown by Szwed in [5] that this mechanism can lead to significant
asymmetries at the beam energy of the order of
$10-20\,$GeV\footnote{$^{\dag}$}{In Ref.~[5] the factorization-breaking
recombination model of fragmentation was used.  In that approach, both the
quark
and the antiquark forming the produced meson originate from the projectile
proton.  They scatter independently but each of them feels different Coulomb
field.}.  Here, if the flavour of $q$ (and also $\vPol_q$) is defined, the
final
asymmetry does not depend on whether $h$ is a PS or a V meson.  This, at high
$x_F$, means that the asymmetries of {\it e.g.\ }$\pi^+$ and $\rho^+$ are close
to each other since both mesons origin from fragmentation of a $u$ quark
polarized equally in both cases.

\noindent
In the reality, one can expect a mixture of the two effects with the second one
vanishing at a sufficiently high energy.  We shall concentrate on the case
{\it a)} and calculate the asymmetry of vector mesons therein.

\medskip
\bigskip
\noindent
{\bf Asymmetry of pions}

\smallskip
\noindent
For the production of pions, it was assumed in [2] that the
cross-section for the reaction (1) can be divided into three parts:

$$ {d\sigma_\uparrow^{p\to h} \over dx_F d^2\vpT} = \left[
{d\sigma_\uparrow^{p\to h} \over dx_F d^2\vpT}
\right]_{\rm quark}^{\rm rank=1} + \left[ {d\sigma^{p\to h} \over dx_F d^2\vpT}
\right]_{\rm quark}^{ {\rm rank}\, \ge\, 2} + \left[ {d\sigma^{p\to h} \over
dx_F
d^2\vpT} \right]_{\rm diquark}^{ {\rm rank}\, \ge\, 2}
\eqno(5)$$
with the successive terms being contributions from observing a leading hadron
of
a string spanned by a quark coming from the projectile, a nonleading one of
that string and a nonleading particle of the string spanned by the remnant
projectile diquark (the leading hadron of that string is a baryon).  The
asymmetry appears in the first term:

$${1 \over \sigma_{\uparrow\rm tot}^{p\to\pi}} \left[
{d\sigma_\uparrow^{p\to\pi} \over dx_F d^2\vpT} \right]_{\rm quark}^{\rm
rank=1}
=
\sum_{q=u,d}\int dx\, dz\, d^2 \vqT\, d^2 \vec{\bar q}_\T
G_q(x, \vqT)\ D_{\pi/q}^{\rm rank=1}(z)
$$
$$\times
{1\over 4}\left(1 - \vPol_q \cdot \vPol_{\bar q}(\vec{\bar q}_\T)\right)\
\rho(\vec{\bar q}_\T)\ \delta(x_F-xz)\ \delta^2(\vpT - \vqT - \vec{\bar q}_\T),
\eqno(6)$$
where $D^{\rm rank=1}$ denotes the part of the fragmentation function
corresponding to the leading meson,

$$
\rho(\vec {\bar q}_\T) = {1 \over \kappa}
                          \exp\left(-\pi {\bar q_\T^2 \over \kappa}\right)
\eqno(7)$$
is the distribution of the transverse momentum of the antiquark $\bar q$ of the
rank-one $q\bar q$ pair produced in the string; $\kappa$ is the string tension.
The production of the subleading hadrons (the second and the third term in
Eq.~(5)) are assumed to be azimuthally symmetric.  Those two remaining terms
have a form similar to Eq.~(6); they have the nonleading parts of the
fragmentation function $D^{{\rm rank}\,\ge\,2}$ and do not have the
polarization
factor $(1 - \vPol_q \cdot \vPol_{\bar q})$.  No parton scattering is included
in Eq~(6).

$\vPol_q$ is the polarization of the leading quark $q$.  The
polarization $\vPol_{\bar q}$ of the antiquark created in the string is
correlated to its transverse momentum $\vec{\bar q}_\T$:

$$ \vPol_{\bar q} = - {2 \bar q_\T (\hat z \times \vec{\bar q}_\T) \over
\kappa + 2 \bar q_\T^2}
\eqno(8)$$
according to the Lund model prescription [4]. This causes the azimuthal
asymmetry which arises in (6) due to the term

$$
{1 \over 4} \, (1 - \vPol_q \cdot \vPol_{\bar q}) \,
\eqno(9)$$
corresponding to the probability that $q$ and $\bar q$ form a spin-singlet
state.

\bigskip
\medskip
\noindent
{\bf Asymmetry of vector mesons}

\smallskip
\noindent
The corresponding probability of forming a spin-1 state is

$$
{3 \over 4} \, (1 + {1 \over 3}\vPol_q \cdot \vPol_{\bar q}) \,.
\eqno(10)$$
Thus, the leading quark contribution to the $\rho$-meson production reads:

$${1 \over \sigma_{\uparrow \rm tot}^{p\to\rho}} \left[
{d\sigma_\uparrow^{p\to\rho} \over dx_F d^2\vpT} \right]_{\rm
quark}^{\rm rank=1} = \sum_{q=u,d}\int dx\, dz\, d^2 \vqT\, d^2 \vec{\bar q}_\T
G_q(x, \vqT)\ D_{\rho/q}^{\rm rank=1}(z)$$

$$\times
{3\over 4}\left(1+{1\over 3}\vPol_q\cdot\vPol_{\bar q}(\vec{\bar q}_\T)\right)\
\rho(\vec{\bar q}_\T)\ \delta(x_F - xz)\ \delta^2(\vpT - \vqT - \vec{\bar
q}_\T)
\eqno(11)$$
One can notice that here the asymetry is opposite in sign and smaller by
the factor $1\over 3$ in magnitude than that of the PS mesons.

The formulae (6) and (11), as they stand, would give the PS to V meson
production ratio equal $1:3$ if the fragmentation functions $D_\pi$ and
$D_\rho$ were identical.  This is due to factors $1\over 4$ and
$3\over 4$ in (9) and (10).  In the string model, the fragmentation functions
for PS and V mesons contain another factors compensating those in (9) and (10)
so that the final PS:V ratio becomes 1:1.  They arise due to the larger mass of
the V mesons.  In any case, they do not influence the asymmetry since they
appear equally in all the three terms in the cross-section (5).

\bigskip
\medskip
\noindent
{\bf Results}

\smallskip
\noindent
In the calculation of the asymmetry we used the same parameters as in Ref.~[2].
We used the quark distribution functions $G_{u/p}(x,\vec q_\T) =
2G_{d/p}(x,\vec
q_\T) = {5\over 2} x^{1/2} (1-x)\rho(\vec q_\T)$; the intrinsic transverse
momentum distribution was assumed to be the same as the tunneling transverse
momentum distribution (7) in the string.  We used the string tension $\kappa =
0.17\,$GeV$^2$ and the flavour-abundance ratio in pair production was taken to
be $u:d:s=3:3:1$.  The used string splitting function was that of the Standard
Lund model, $f(z) = (1+C) (1-z)^C$, with $C=0.3$.  The leading part of the
fragmentation function is $D_{h/q}^{\rm rank=1}(z) = c_h f(z)$, where $c_h$ is
an appropriate flavour factor [2].  As motivated by the results of [2] and
comparison to the data [8,9] we took the maximal possible polarizations of the
$u$ and $d$ valence quarks in a transversely polarized proton:  $\vPol_u = +1$
and $\vPol_d = -1$ at $x=1$.

In Fig.~1 we show the results obtained for the asymmetry $A_N$ of the
$\rho^\pm$
mesons production, at $200\,$GeV beam momentum, compared to that of the charged
pions coming from the Ref.~[2].  The pion data of the E704 collaboration [8]
are
also shown.  The full and the dashed lines correspond to different dependences
of the quark transversity $\vPol_q$ on the momentum fraction $x$.  They were
chosen to be decreasing towards small $x$ as $x^2$ and $x$, respectively.
These
parametrizations gave in [2] the best agreement with $x_F$-dependence of the
data.  The $p_\T$ cuts are the same for $\rho$ mesons as for the pions and
correspond to the experimental ones.  The asymmetry of the $\rho^0$ mesons is
shown in Fig.~2 also compared to that of $\pi^0$ and to the data [9].  The
asymmetries are here smaller than in the case of the charged mesons.
Nevertheless, a measurement of $\rho_0$ can be easier.

As already argued, the ratio of the asymmetries

$$
R_{\rho/\pi} = {A_N^\rho \over A_N^\pi}
\eqno(12)$$
is equal $-{1\over 3}$.  It would be of large interest to verify
this prediction experimentally.

This result is more general than the used model. One can expect such a ratio in
any calculation where the asymmetry arises in fragmentation due to a
correlation between the spin and the transverse momentum of the antiquark
accompanying the leading quark to form the meson and where the nonrelativistic
quark model is assumed for this meson.

\bigskip
\medskip
\noindent
{\bf Asymmetry of parton scattering}

\smallskip
\noindent
Violating the rule (12) can be an indication of appearing of the Szwed effect
[5,6] {\it i.e.\ }significant asymmetry in parton scattering.  In that model,
originally constructed to describe the polarization of hyperons $\Lambda$ [6],
the asymmetry of a transversely polarized quark scattered on a Coulomb-like
field is given by:

$$ \Aqq = 2 C_S \alpha_S
       {m k \sin^3(\theta/2) \ln[\sin(\theta/2)]
       \over
       [m^2 + k^2 \cos^2(\theta/2)] \cos(\theta/2) }
       {\vec k \times \vec k' \over |\vec k \times \vec k' |}
\eqno(13)$$
where $m$, $\vec k$ and $\vec k'$ are the mass, the initial and the final
momentum of the scattered quark and $k = |\vec k|$.  $\theta$ is the scattering
angle in the frame where $|\vec k| = |\vec k'|$.  $C_S$ is the constant
characterizing the external strong field.  The sign of the asymmetry depends on
the sign of the constant $C_S$ in (13) or on whether the field source is
``quark-like'' or ``antiquark-like''.  In the first case $C_S$ is positive and
the asymmetry $\Aqq$ negative, in the latter $C_S$ is negative and $\Aqq$
positive.  It would be interesting to see what the asymmetry is in quark--gluon
scattering.  The formula (13) comes from a second-order calculation in QED and
can only have intuitive meaning for the strong interactions.

Hence, if the asymmetry of quark scattering is treated as a correction to that
of fragmentation, $R_{\rho/\pi} < -{1\over 3}$ will mean negative $\Aqq$ or
scattering off a ``quark-like'' field and $-{1\over 3} < R_{\rho/\pi} < 0$ will
mean positive $\Aqq$ and ``antiquark-like'' field.  As explicitly seen from
Eq.~(13), at sufficiently high energy ($k\gg m$) the asymmetry of quark
scattering vanishes.  The energy scale where it appears can be an interesting
hint about the scale of the masses of partons being scattered in a $pp$
collision.

$R_{\rho/\pi}\ne -{1\over 3}$ could also mean a violation of the
nonrelativistic
quark model, from which formulae (9) and (10) come form. If this were the
case, then the high-energy limit of $R_{\rho/\pi}$ could be taken as the
reference value, instead of $-{1\over 3}$ and the above analysis could be also
made.

\bigskip
\medskip
\noindent
{\bf Conclusions}

\smallskip
\noindent
To summarize, we have calculated the single spin asymmetry of $\rho$ meson
production in $p\up p$ collisions in the framework of Ref.~[2], where the
asymmetry of pions has been obtained.  The asymmetry was generated only in the
fragmentation function.  The two asymmetries are found to by opposite in sign,
the ratio of that of $\rho$ to that of $\pi$ being -1/3.  Violating this rule
can indicate a significant contribution from the asymmetry of the transversely
polarized parton subprocess.

However, one should note that the proposed measurement would be possible
only if the asymmetry of $\rho$ mesons were compared to that of directly
produced pions (not coming from decays of V mesons). If no such cuts were made
(as is the case in the E704 data) the method would have to be modified.
Another important constraint is the fact that the discussed model is a good
approximation only in the region of high $x_F$. For smaller $x_F$ it
includes more assumptions and approximations. For example, the asymmetry
of higher-rank hadrons has not been taken into account.

\bigskip
\medskip
\noindent
{\bf Acknowledgements}
\smallskip

\noindent
I am very grateful to Xavier Artru for many enlightening discussions and
suggestions about improving the manuscript.  Financial support from the
IN2P3--Poland scientific exchange programme and from the Polish Government
grants of KBN no.\ 2~0054~91~01, 2~0092~91~01 and 2~2376~91~02 are also
acknowledged.

\bigskip
\medskip
\noindent
{\bf References}
\smallskip

\item{[1]} {\it See e.g.\ }J.C.~Collins, in Proceedings of PANIC XIII ---
Particles and Nuclei International Conference at Perugia, Italy, 28th June --
2nd July 1993 (World Scientific, Singapore, 1993) {\it and references quoted
therein}

\item{[2]} X.~Artru, J.~Czy\.zewski, H.~Yabuki, {\it IPN Lyon preprint} {\bf
LYCEN/9423}, {\it Cracow JU preprint} {\bf TPJU 12/94}, hep-ph/9405426

\item{[3]} J.C.~Collins, \NP {\bf B396}, 161 (1993); J.C.~Collins,
S.F.~Heppelmann and G.A.~Ladinsky, \NP {\bf B420}, 565 (1994)

\item{[4]} B.~Andersson, G.~Gustafson, G.~Ingelman, T.~Sj\"ostrand, \PRep\ {\bf
97} 31 (1983)

\item{[5]} J.~Szwed, Proceedings of the 9$^{\rm th}$ International Symposium
``High Energy Spin Physics'' held at Bonn, 6--15 Sep.\ 1990, Springer Verlag
1991;

\item{[6]} J.~Szwed, \PL {\bf B105}, 403 (1981)

\item{[7]} J.P.~Ralston and D.E.~Soper, \NP {\bf B152}, 109 (1979); X.~Artru
and M.~Mekhfi, \ZP {\bf C45}, 669 (1990); J.L.~Cortes, B.~Pire and
J.P.~Ralston,
\ZP {\bf C55}, 409 (1992); R.L.~Jaffe and Xiangdong Ji, \NP {\bf B375}, 527
(1992)

\item{[8]} D.L.~Adams {\it et al.}, \PL {\bf B264}, 462 (1991)

\item{[9]} D.L.~Adams {\it et al.}, \PL {\bf B261}, 201 (1991); \ZP {\bf C56},
181 (1992)

\bigskip
\medskip
\noindent
{\bf Figure captions}

\smallskip

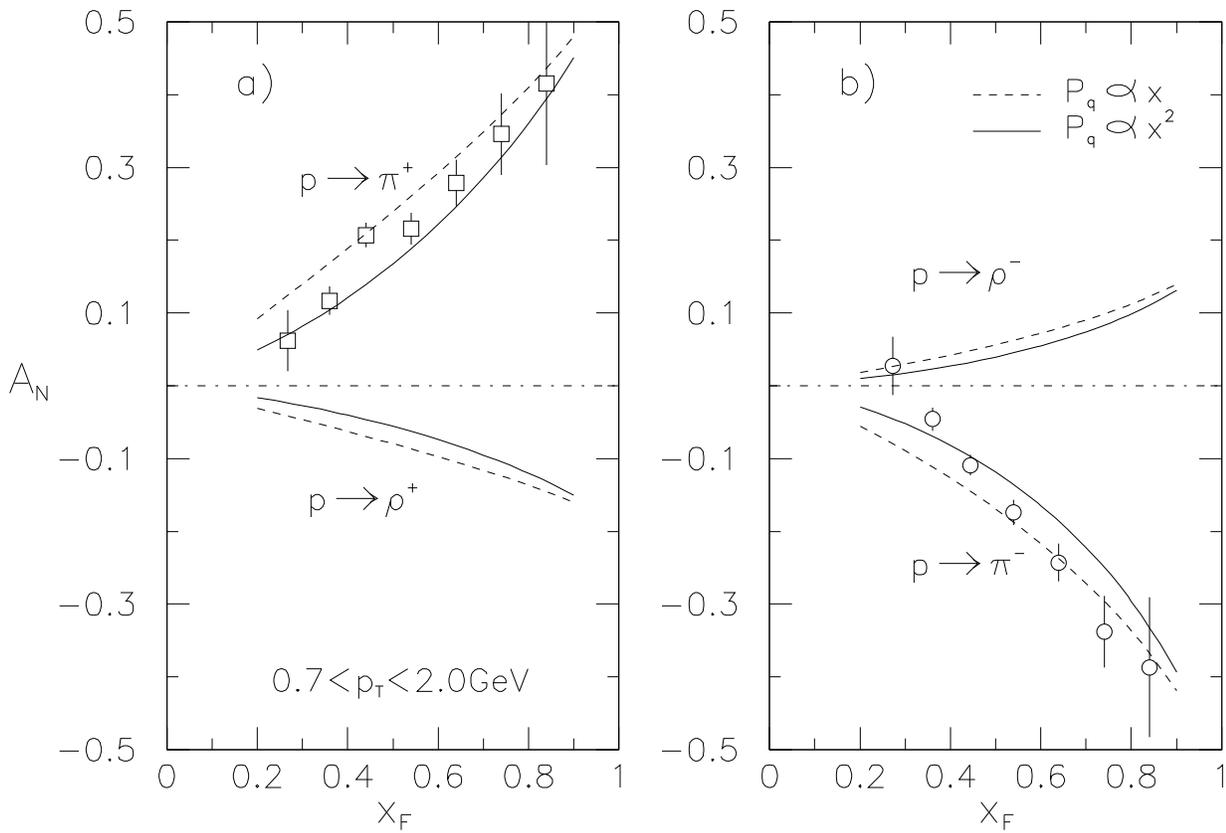
\item{Fig.~1} Single spin asymmetry of charged $\rho$ mesons compared to that
of
charged pions and the E704 data [8].

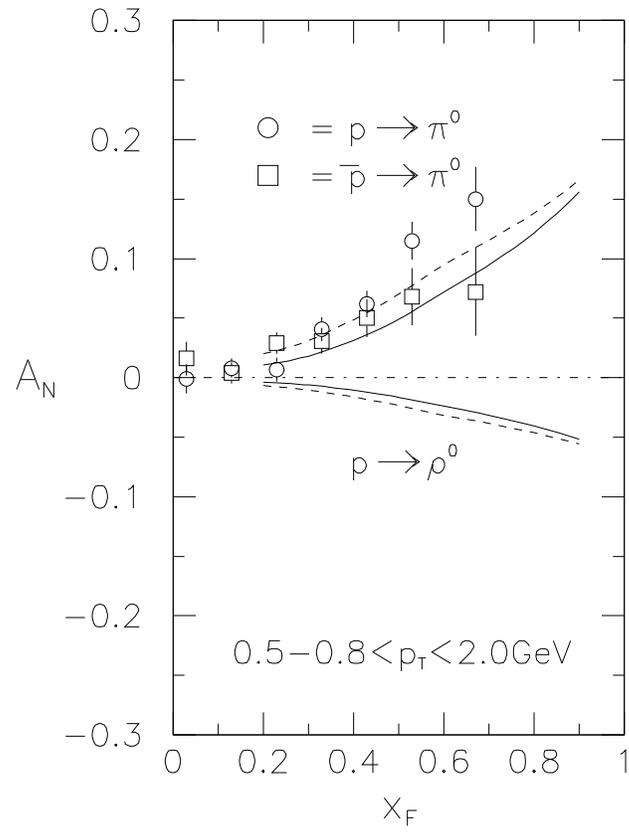
\item{Fig.~2} Asymmetries of neutral $\rho$ and $\pi$. The $\pi^0$ asymmetry
data are from [9].

\end